\documentclass[twocolumn,A4]{article}
\usepackage[dvips]{graphics}

\begin{document}

\onecolumn

\begin{center}
{\bf{\Large Anomalous Quantum Diffusion in Order-Disorder Separated Double 
Quantum Ring}}\\
~\\
Santanu K. Maiti$^{\dag,\ddag,*}$ \\
~\\
{\em $^{\dag}$Theoretical Condensed Matter Physics Division, 
Saha Institute of Nuclear Physics, \\
1/AF, Bidhannagar, Kolkata-700 064, India \\
$^{\ddag}$Department of Physics, Narasinha Dutt College,
129, Belilious Road, Howrah-711 101, India} \\
~\\
{\bf Abstract}
\end{center}
A novel feature for control of carrier mobility is explored in an
order-disorder separated double quantum ring, where the two rings thread 
different magnetic fluxes. Here we use simple tight-binding formulation 
to describe the system. In our model, the two rings are connected through 
a single bond and one of the rings is subjected to impurity, keeping the 
other ring as impurity free. In the strong impurity regime, the electron 
diffusion length increases with the increase of the impurity strength, 
while it decreases in the weak impurity regime. This phenomenon is 
completely opposite to that of a conventional disordered double quantum 
ring, where the electron diffusion length always decreases with the 
increase of the disorder strength. 
\vskip 1cm
\begin{flushleft}
{\bf PACS No.}: 73.23.Ra; 73.23.-b; 73.63.-b \\
~\\
{\bf Keywords}: Persistent current; Drude weight; Double quantum ring; 
Impurity.
\end{flushleft}
\vskip 4.5in
\noindent
{\bf ~$^*$Corresponding Author}: Santanu K. Maiti

Electronic mail: santanu.maiti@saha.ac.in
\newpage
\twocolumn

\section{Introduction}

Over the last few decades, the physics at sub-micron length scale provides
enormous evaluation both in terms of our understanding of basic physics
as well as in terms of the development of revolutionary technologies.
In this length scale, the so-called mesoscopic or nanoscopic regime,
several characteristic quantum length scales for the electrons such as
system size and phase coherence length or elastic mean free path and
phase coherence length are comparable. Due to the dominance of the quantum
effects in the mesoscopic/nanoscopic regime, intense research in this 
field has revolved its richness. The most significant issue is probably
the persistent currents in small normal metal rings. In thermodynamic 
equilibrium, a small metallic ring threaded by magnetic flux $\phi$ 
supports a current that does not decay dissipatively even at non-zero 
temperature. It is the well-known phenomenon of persistent current in 
mesoscopic normal metal rings which is a purely quantum mechanical effect 
and gives an obvious demonstration of the Aharonov-Bohm effect.$^1$ 
The possibility of persistent current was predicted in the very early days 
of quantum mechanics by Hund,$^2$ but their experimental evidences came 
much later only after realization of the mesoscopic systems. In $1983$, 
B\"{uttiker} {\em et al.}$^3$ predicted theoretically that persistent 
current can exist in mesoscopic normal metal rings threaded by a magnetic 
flux $\phi$, even in the presence of impurity. In a pioneering experiment, 
Levy {\em et al.}$^4$ first gave the experimental evidence of persistent 
current in the mesoscopic normal metal ring, and later, the existence of 
the persistent current was further confirmed by several experiments.$^{5-8}$ 
Though the phenomenon of persistent current has been addressed quite 
extensively over the last twenty years both theoretically$^{9-27}$
as well as experimentally,$^{4-8}$ but yet we cannot resolve the
controversy between the theory and experiment. The main controversies come 
in the determinations of (a) the current amplitude, (b) flux-quantum 
periodicities, (c) low-field magnetic susceptibilities, etc. In recent 
works,$^{24-26}$ we have pointed out that the higher order hopping integrals, 
in addition to the nearest-neighbor hopping integral, have a significant 
role to enhance the current amplitude (even an order of magnitude). 
In other recent work,$^{27}$ we have focused that the low-field magnetic 
susceptibility can be predicted exactly only for the one-channel systems 
with fixed number of electrons, while for all other cases it becomes random. 
To grasp the experimental behavior of the persistent current, one 
has to focus attention on the interplay of quantum phase coherence, 
disorder and electron-electron correlation and this is a highly
complex problem.

Using the advanced molecular beam epitaxial growth technique, one can 
easily fabricate a quantum system where the impurities are located only 
in some particular region of the system, keeping the other region free 
from any impurity. This is completely opposite from a conventional 
disordered system, where the disorders are given uniformly throughout 
the system. Traditional wisdom is that, the larger the disorder stronger 
the localization.$^{28}$ However, some recent experimental 
studies$^{29-31}$ as well as theoretical investigations$^{32-35}$ on these 
special class of systems where the disorders are not distributed uniformly,
have yielded completely different behavior which predicts that the electron 
diffusion length decreases in the weak disorder regime, while it increases 
in the strong disorder regime. Motivated with these results, in this article,
we focus our attention in an order-disorder separated double quantum ring 
system. To reveal the variation of the electron diffusion length in such 
a particular system, here we study the behavior of persistent current and 
Drude weight and our results may illuminate some of the unusual 
experimental results for such diverse transport property. The parameter 
Drude weight $D$ characterizes the conducting nature of the system as 
originally introduced by Kohn.$^{36}$ 
In our present model, two mesoscopic rings, threaded by different magnetic 
fluxes, are connected by a single bond and impurities are given in 
any one of these two rings, while the other ring becomes impurity free. 
For this order-disorder separated double quantum ring, we observe an 
anomalous behavior of electron mobility in which the electron diffusion
length increases with the increase of the impurity strength in the 
strong impurity regime, while the diffusion length decreases in the weak 
impurity regime. This phenomenon is completely opposite to that of a 
conventional disordered double quantum ring, in which the electron diffusion
length always decreases with the increase of the disorder strength. 

In what follows, we describe the model and the method in Section $2$. 
Section $3$ contains the significant results and the discussion, and 
finally, we summarize our results in Section $4$.

\section{The model and the method}

\begin{figure}[ht]
{\centering \resizebox*{6.5cm}{3.2cm}{\includegraphics{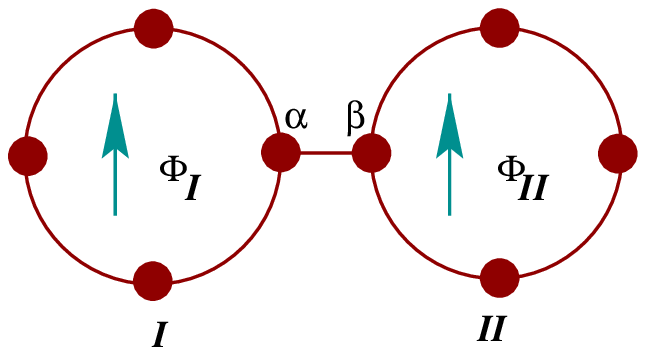}}\par}
\caption{Schematic view of a double quantum ring in which the two rings 
thread magnetic fluxes $\phi_I$ and $\phi_{II}$ respectively. These two 
rings are connected through the lattice sites $\alpha$ and $\beta$. 
The filled circles correspond to the position of the atomic sites (for 
color illustration, see the web version).}
\label{ring}
\end{figure}

The schematic representation of a double quantum ring is shown in 
Fig.~\ref{ring} where the two rings, threaded by different magnetic fluxes,
are connected by a single bond. In the non-interacting picture, the system 
is usually modeled by a single-band tight-binding Hamiltonian,
\begin{eqnarray}
H & = & \sum_i \epsilon_i^{I} c_i^{\dagger} c_i + v_I \sum_{<ij>}
\left[e^{i \theta_I} c_i^{\dagger} c_j+ e^{-i \theta_I} c_j^{\dagger} 
c_i \right] \nonumber \\ 
& + & \sum_k \epsilon_k^{II} c_k^{\dagger} c_k + v_{II} \sum_{<kl>}
\left[e^{i \theta_{II}} c_k^{\dagger} c_l+ e^{-i \theta_{II}} c_l^{\dagger} 
c_k \right] \nonumber \\ 
& + &  v_{\alpha\beta}
\left[c_{\alpha}^{\dagger}c_{\beta}+c_{\beta}^{\dagger}c_{\alpha}\right]
\label{hamil1}
\end{eqnarray}
Here $\epsilon_i^I$'s ($\epsilon_i^{II}$'s) are the site energies in the
ring $I$ (ring $II$), $c_i^{\dagger}$ ($c_k^{\dagger}$) is the creation 
operator of an electron at site $i$ ($k$) of the ring $I$ (ring $II$) and
$c_i$ ($c_k$) is the annihilation operator of an electron at site $i$ ($k$)
of the ring $I$ (ring $II$), $v_I$ ($v_{II}$) is the hopping strength 
between nearest-neighbor sites in the ring $I$ (ring $II$), and 
$v_{\alpha\beta}$ gives the hopping strength between these two rings. 
In this expression, $\theta_I=2\pi \phi_I/N_I$ and $\theta_{II}=2\pi 
\phi_{II}/N_{II}$ are the phase factors due to the fluxes $\phi_I$ and 
$\phi_{II}$ (measured in units of $\phi_0=ch/e$, the elementary flux 
quantum), respectively where $N_I$ and $N_{II}$ correspond to the total 
number of atomic sites in the ring $I$ and ring $II$, respectively. 
In order to introduce the impurities in the system, we choose the site 
energies ($\epsilon_i$'s, omitting the ring index in the superscript) 
from the relation: $\epsilon_i=W \cos(i\lambda \pi)$, where $W$ is the 
strength of the disorder and $\lambda$ is an 
irrational number, and as a typical example we take it as the golden 
mean $\left(1+\sqrt{5}\right)/2$. Setting $\lambda=0$, we get back the 
pure system with identical site potential $W$. The idea of considering 
such an incommensurate potential is that, for such a correlated disorder 
we do not require any configuration averaging and therefore the numerical 
calculations can be done in the low cost of time. Now to achieve the 
order-disorder separated double quantum ring, we introduce the correlated 
disorder in any one of the rings, keeping the other one as impurity free.

At absolute zero temperature, the persistent currents in the two rings
can be calculated from the expressions, 
\begin{eqnarray}
I(\phi_I) = - \frac{\partial{E(\phi_I, \phi_{II})}}{\partial{\phi_I}} \\
I(\phi_{II}) = - \frac{\partial{E(\phi_I, \phi_{II})}}{\partial{\phi_{II}}}
\label{curr}
\end{eqnarray}
where, $I(\phi_I)$ and $I(\phi_{II})$ correspond to the currents in the
ring $I$ and ring $II$, respectively and $E(\phi_I, \phi_{II})$ represents
the ground state energy of the complete system. We evaluate this energy 
exactly to understand unambiguously the anomalous behavior of persistent 
current, and this is achieved by exact diagonalization of the tight-binding 
Hamiltonian Eq.~(\ref{hamil1}).

Now the response of the double quantum ring system to a uniform 
time-dependent electric field can be determined in terms of the Drude weight 
$D$,$^{37-38}$ a closely related parameter that characterizes the conducting 
nature of the system as originally noted by Kohn.$^{36}$ The Drude weights 
for the two rings can be calculated through the relations,$^{39}$
\begin{eqnarray}
D_I=\left . \frac{N_I}{4\pi^2}\left(\frac{\partial{^2E(\phi_I, \phi_{II})}}
{\partial{\phi_I}^{2}}\right) \right|_{\phi_I \rightarrow 0,
\phi_{II} \rightarrow 0} \\
D_{II}=\left . \frac{N_{II}}{4\pi^2}\left(\frac{\partial{^2E(\phi_I, 
\phi_{II})}} {\partial{\phi_{II}}^{2}}\right) \right|_{\phi_I \rightarrow 0,
\phi_{II} \rightarrow 0} 
\label{drude}
\end{eqnarray}
where $D_I$ and $D_{II}$ represent the Drude weights for the ring $I$ and 
ring $II$ respectively.

Our main aim in this article is the determination of the conducting 
properties of an order-disorder separated double quantum ring, which can
be computed through the parameters $D_I$ and $D_{II}$. From these 
parameters we can clearly describe the mobility of the charge carriers
in the system and accordingly, the variation of the electron diffusion
length might be expected.

\section{Results and discussion}

In the order-disorder separated double quantum ring, we introduce the 
correlated disorder in ring $I$ (for the sake of simplicity), keeping the 
ring $II$ as impurity free. Throughout the numerical computations, we take 
the values of the different parameters as: $v=-1$, $v_{\alpha\beta}=-1$ 
and for the sake of simplicity, we use the units where $c=1$, $e=1$ and 
$h=1$. During these calculations, we fix the chemical potential ($\mu$) 
for all the systems to a constant value $0$. The main focus of this article 
is to describe how
the disordered states affect the ordered states in the order-disorder
separated system. Since for such a system the impurities are introduced
in the ring $I$, we evaluate the conducting properties of the double 
quantum ring by measuring the Drude weight $D_{II}$. This actually provides
the response of the ordered states in presence of the disordered states.
\begin{figure}[ht]
{\centering \resizebox*{8cm}{9cm}{\includegraphics{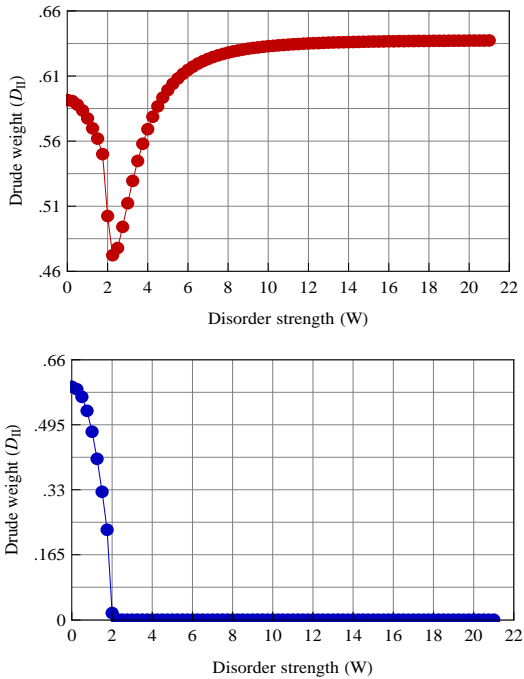}}\par}
\caption{Drude weight ($D_{II}$) of the ring $II$ as a function of the 
disorder strength ($W$) for the systems with $N_I=30$, $N_{II}=30$ and 
the fixed chemical potential $\mu=0$. The red and the blue lines 
correspond to the order-disorder separated and the complete disordered 
double quantum ring systems respectively (for color illustration, see
the web version).}
\label{drude1}
\end{figure}
Otherwise, if we measure the parameter $D_I$, then we will get the trivial 
result as obtained in a traditional disordered system since the response
of the disordered states will not be changed by coupling these states 
with the ordered states. 

In Fig.~\ref{drude1}, we show the variation of the Drude weight $D_{II}$
as a function of the disorder strength $W$ for some double quantum rings,
where we choose $N_I=30$ and $N_{II}=30$. The chemical potential $\mu$
is fixed to $0$. The red and the blue curves represent the results for the
order-disorder separated and the complete disordered double quantum rings,
respectively. From the results it is observed that, in the complete 
disordered double quantum ring the Drude weight sharply decreases with 
the increase of the disorder strength and eventually it drops to zero. 
Therefore, we can say that for such a system the electron diffusion length 
as well as the electron mobility decreases sharply with the disorder 
strength. Such a behavior can be well understood from the theory of Anderson 
localization, where we get more localization with the increase of the 
disorder strength.$^{28}$ The 
anomalous behavior is observed when the impurities are given only in any 
one of the two rings, keeping the other one as impurity free i.e., for the 
order-disorder separated system. Our results predict that the Drude weight
initially decreases with the increase of the disorder strength, but after 
reaching to a minimum it again increases with the strength of the disorder. 
Such a phenomenon is completely opposite to that of the traditional 
disordered system and can be justified in the following way. For the 
order-disorder separated double quantum ring, the energy spectra of 
the disordered ring are gradually separated from the energy spectra of the 
ordered ring with the increase of the disorder strength $W$. Therefore,
the influence of random scattering in the ordered ring due to the strong
localization in the disordered ring decreases. It has been examined that 
the energy spectrum of the order-disorder separated double quantum ring
with large disorder contains localized tail states with much small and
central states with much large values of localization length, contributed
approximately by disordered and ordered rings, respectively. Hence the 
central states gradually separated from the tail states and delocalized 
with the increase of the strength of the disorder. Thus we see that, for 
the coupled order-disorder separated double quantum ring, the coupling 
between the localized states with the extended states is strongly influenced 
by the strength of the disorder, and this coupling is inversely proportional 
to the disorder strength $W$. Accordingly, in the limit of weak disorder the 
coupling effect is significantly high, while the coupling effect becomes 
very weak in the strong disorder regime. Hence, in the limit of weak disorder 
the electron transport is strongly influenced by the impurities at the 
disordered ring such that the electron states are scattered more and 
therefore the electron diffusion length decreases which manifests the lesser
electron mobility. On the other hand, for the stronger disorder limit the
extended states are weakly influenced by the disordered ring and the coupling
effect gradually decreases with the increase of the disorder strength which
provides the larger electron mobility in the strong disorder limit. This 
reveals that the electron diffusion length increases in this limit. For large
enough impurity strength, the extended states are almost unaffected by the
impurities at the disordered ring and in that case the electrons are carried
only by these extended states in the ordered ring which is the trivial limit.
So the novel phenomenon will be observed only in the intermediate limit 
of $W$.

\begin{figure}[ht]
{\centering \resizebox*{8cm}{9cm}{\includegraphics{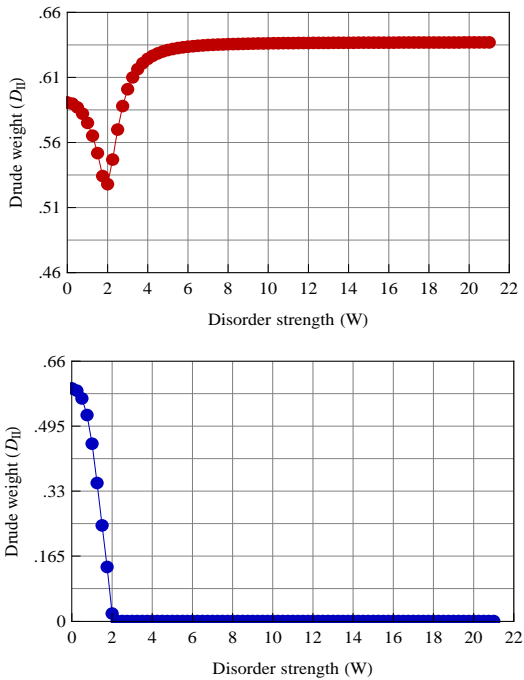}}\par}
\caption{Drude weight ($D_{II}$) of the ring $II$ as a function of the 
disorder strength ($W$) for the systems with $N_I=50$, $N_{II}=50$ and 
the chemical potential $\mu=0$. The red and the blue lines correspond to 
the order-disorder separated and the complete disordered double quantum 
ring systems respectively (for color illustration, see the web version).}
\label{drude2}
\end{figure}

In order to emphasize the dependence of the electron mobility on the system
size, here we focus our attention on the results those are plotted in 
Fig.~\ref{drude2}. In this figure, we display the Drude weight for some 
typical double quantum rings, where we fix $N_I=50$ and $N_{II}=50$. 
Similar to the previous systems, here we also take $\mu=0$ for these 
systems. The red and the blue lines correspond to the same meaning as in 
Fig.~\ref{drude1}. From this figure 
(Fig.~\ref{drude2}) it is also observed that, the Drude weight in the 
order-disorder separated double quantum ring decreases with the increase 
of the disorder strength $W$ in the weak disorder regime, while it 
increases with the strength $W$ in the strong disorder regime. On the 
other hand, the Drude weight always decreases with the strength of the 
disorder for the complete disordered system, as expected. Though the 
results plotted in Fig.~\ref{drude2} seem to be quite similar in nature 
with the results those are described in Fig.~\ref{drude1}, but the 
significant point is that, the typical magnitude of the Drude weight 
strongly depends on the size of both these two rings which manifest the 
finite quantum size effects. Now the other significant factor that raises
to our mind is the existence of the location of the minimum in the Drude 
weight versus disorder curves of the order-disorder separated double 
quantum rings.
This minimum can be implemented as follows. The carrier mobility in the 
order-disorder separated double quantum ring is controlled by the two 
competing mechanisms. One is the random scattering in the ordered ring 
due to the localization in the disordered ring which tends to decrease 
the carrier mobility, and the other one is the vanishing influence of 
random scattering in the ordered ring due to the strong localization in 
the disordered ring which provides the enhancement of the carrier mobility.
Depending on the ratio of the total number of atomic sites in the 
disordered ring to the total number of atomic sites in the ordered ring,
the vanishing effect of random scattering from the ordered states dominates 
over the non-vanishing effect of random scattering from these states for 
a particular disorder strength $(W=W_c)$, which provides the location of 
the minimum in the Drude weight versus disorder curve.

\section{Concluding remarks}

In conclusion, we have established a novel feature for control of the 
electron diffusion length in an order-disorder separated double quantum 
ring in which the two rings thread different magnetic fluxes. From our 
study it has been observed that, in the order-disorder separated double 
quantum ring, the electron diffusion length increases with the increase of the 
disorder strength in the strong disorder regime, while it decreases in the 
weak disorder regime. Such a peculiar behavior is completely opposite to that 
of the conventional disordered systems, where the electron diffusion length 
always decreases with the increase of the disorder strength. Lastly, we have 
noticed that, both the electron mobility and the location of the minimum in
the Drude weight versus disorder curve strongly depend on the size of both 
the two rings which manifest the finite quantum size effects. Our theoretical 
results in this article might be helpful to illuminate some of the unusual 
experimental phenomena which have been observed in the order-disorder 
separated quantum systems.$^{29-31}$

Throughout our study, we have ignored the effect of the electron-electron 
(e-e) correlation since the inclusion of the e-e correlation will not 
provide any new significant result in our present investigations. 

\vskip 0.3in
\noindent
{\bf\Large Acknowledgment}
\vskip 0.2in
\noindent
I acknowledge with deep sense of gratitude the illuminating comments and
suggestions I have received from Prof. S. Sil during the calculations.

\end{document}